\documentclass[prd,aps,twocolumn,nofootinbib,preprintnumbers,superscriptaddress,preprintnumbers,balancelastpage,longbibliography]{revtex4-2}
\usepackage[english]{babel}

\usepackage{amsmath,mathtools,physics,xfrac}
\usepackage[normalem]{ulem}
\usepackage{graphicx}
\usepackage{afterpage}
\usepackage{float}
\usepackage{subfigure}
\usepackage{rotating}
\usepackage{multirow}
\usepackage{tabularx}
\usepackage{booktabs}
\usepackage{fancyhdr}
\usepackage[utf8]{inputenc}
\usepackage{theorem}
\usepackage{moreverb}
\usepackage{euscript}
\usepackage{psfrag}
\usepackage{slashed}
\usepackage{mathtools}
\usepackage{makecell}
\usepackage{adjustbox}
\usepackage{dcolumn}
\usepackage{bm}
\usepackage[dvipsnames]{xcolor}
\usepackage{graphics}
\usepackage{hyperref,url}

\hypersetup{
     colorlinks   = true,
     citecolor    = Blue,
     urlcolor     = Blue,
     linkcolor    = Blue
}

\usepackage{color,xcolor}

\definecolor{darkblue}{rgb}{0.0,0.0,0.75}
\definecolor{darkred}{rgb}{0.6,0.0,0}
\definecolor{darkgreen}{rgb}{0.0,0.6,0.}

\usepackage{hhline, array, makecell}
\usepackage{aas_macros}
\usepackage{orcidlink}

\hypersetup{
     colorlinks   = true,
     citecolor    = blue,
     urlcolor     = blue,
     linkcolor    = blue
}

\usepackage{color,xcolor}

\definecolor{darkblue}{rgb}{0.0,0.0,0.75}
\definecolor{darkred}{rgb}{0.6,0.0,0}
\definecolor{darkgreen}{rgb}{0.0,0.6,0.}
\newcommand{\AD}[1]{{\color{red} [AD: #1]}}

\newcommand{\K}{\,\mathrm{K}}

\newcommand{\mum}{\,\mu\mathrm{m}}

\newcommand{\eV}{\,\mathrm{eV}}

\newcommand{\MeV}{\,\mathrm{MeV}}
\newcommand{\keV}{\,\mathrm{keV}}
\newcommand{\meV}{\,\mathrm{meV}}

\newcommand{\cmsq}{\,\mathrm{cm^{2}}}
\newcommand{\kms}{\,\mathrm{km\,s^{-1}}}

\newcommand{\mchi}{m_\chi}

\newcommand{\mphi}{m_\phi}
\newcommand{\bq}{\boldsymbol{q}}

\newcommand{\bv}{\boldsymbol{v}}
\newcommand{\fmed}{F_\mathrm{med}}

\newcommand{\schie}{\sigma_{\chi e}}
\newcommand{\muce}{\mu_{\chi e}}

\begin{document}
\title{Sub-MeV Dark Matter Detection with Bilayer Graphene}

\author{Anirban Das\,\orcidlink{0000-0002-7880-9454}}
\email{anirbandas@snu.ac.kr}
\affiliation{Department of Physics \& Astronomy, Seoul National University, Seoul 08826, Korea}

\author{Jiho Jang\,\orcidlink{0000-0002-0687-891X}}
\email{jiho.jang@snu.ac.kr}
\affiliation{Department of Physics \& Astronomy, Seoul National University, Seoul 08826, Korea}

\author{Hongki Min\,\orcidlink{0000-0001-5043-2432}}
\email{hmin@snu.ac.kr}
\affiliation{Department of Physics \& Astronomy, Seoul National University, Seoul 08826, Korea}

\affiliation{Center for Theoretical Physics, Seoul National University, Seoul 08826, Korea}

\begin{abstract}
The light dark matter mass regime has emerged as the next frontier in the direct detection experiment due to the lack of any detection signal in the higher mass range.
In this paper, we propose a new detector material, a bilayer stack of graphene to detect sub-MeV dark matter. Its voltage-tunable low energy sub-eV electronic band gap makes it an excellent choice for the detector material of a light dark matter search experiment. We compute its dielectric function using the random phase approximation and estimate the projected sensitivity for sub-MeV dark matter-electron scattering and sub-eV dark matter absorption. We show that a bilayer graphene dark matter detector can have competitive sensitivity as other candidate target materials, like a superconductor, but with a tunable threshold energy in this mass regime. The dark matter scattering rate in bilayer graphene is also characterized by a daily modulation from the rotation of the Earth which may help us mitigate the backgrounds in a future experiment. We also outline a detector design concept and provide noise estimates that can be followed to setup an experiment in future.
\end{abstract}


\maketitle

\section{Introduction}\label{sec:intro}


The search for sub-MeV particle dark matter (DM) has become the next frontier in the direct detection experiments. Most of the direct detection experiments in the last few decades have concentrated in heavier WIMP-like DM regime above a GeV, but have failed to yield any conclusive signal yet. On the other hand, the lack of experiments in the lighter mass regime have left a vast parameter space unexplored till date. However, particle DM with sub-MeV mass can be produced with the observed relic abundance in several theoretical scenarios, such as strongly interacting DM\,\cite{Hochberg:2014dra}, freeze-in or freeze-out DM\,\cite{Boehm:2003ha,Boehm:2003hm,Fayet:2004bw,Izaguirre:2015yja,Chang:2019xva}, and other nonstandard DM scenarios\,\cite{DAgnolo:2018wcn}. The direct detection experiments looking for heavier DM use nuclear and electronic recoils or excitations in the detector from DM scattering, and have larger energy threshold typically above $\mathcal{O}(10)\eV$. However, DM being nonrelativistic with a typical velocity $240\kms$ in the Milky Way halo, its kinetic energy is about six orders of magnitude smaller than its mass. As a result, those experiments lose sensitivity for lighter sub-MeV DM scattering as the energy transfer $\lesssim\mathcal{O}(\mathrm{few})\eV$ falls short of their threshold. 

Therefore new ideas are needed for light DM search. In the past few years, several target materials and experimental techniques have been suggested for light DM search using electronic excitations of lower energy some of which have already been realized in experiments. A few examples are athermal phonon sensors made of semiconductors\,\cite{CRESST:2017ues,CRESST:2019jnq,SuperCDMS:2020aus,Ren:2020gaq,CPD:2020xvi,tesseract}, single charge detectors\,\cite{SENSEI:2020dpa,DAMIC-M:2023gxo,Oscura:2022vmi}, quantum defects in diamond and silicon carbide\,\cite{Rajendran:2017ynw,Marshall:2020azl}, superfluid helium\,\cite{Hertel:2018aal,Maris:2017xvi,Anthony-Petersen:2023ykl}, magnetic microcalorimeters\,\cite{Kim:2020gni}, superconducting devices\,\cite{Hochberg:2015pha,Hochberg:2019cyy,Chiles:2021gxk,Kim_2022,Das:2022srn,Fink:2023tvb,Das:2023zcv}, Migdal effect in semiconductor\,\cite{Liang:2022xbu}, doped semiconductor\,\cite{Du:2022dxf}, three-dimensional Dirac materials\,\cite{Hochberg:2017wce,Geilhufe:2019ndy,Huang:2023nms}, quantum dots\,\cite{Blanco:2022cel} etc. We encourage the reader to refer to Ref.\,\cite{Essig:2022dfa} for a detailed compilation of such ideas. While these studies have broadened the future outlook of light DM search, each detection technology comes with its unique signal readout and background challenges that are needed to be overcome. Also, some of the experiments have been seeing a background signals in the lower energy channels below a few $100\eV$ that is limiting the progress toward achieving lower energy threshold\,\cite{Fuss:2022fxe}. Hence, it is important to explore various novel materials to optimize the future sensitivity in an experiment.

Apart from the halo DM, low threshold devices are also needed to probe Earth-bound thermalized DM population near the surface of the Earth. Such a population of DM is predicted in a class of models where DM interacts relatively strongly with the ordinary matter\,\cite{Gould:1989hm,Banks:2021sba,DeLuca:2018mzn,Dasgupta:2020dik,Pospelov:2020ktu,Leane:2022hkk}. In such a scenario, a fraction of the DM coming from the Milky Way halo will scatter multiple times in the Earth's crust, lose energy, and get gravitationally captured. These DM particles will thermalize with the local temperature which is about $300\K$ near the Earth's surface. This translates to about $26\meV$ for the typical kinetic energy of a DM particle. Therefore, detecting this thermalized DM population asks for either detectors with energy threshold below $\lesssim100\meV$, or ways to boosting them to higher energy, or annihilation into Standard Model particles\,\cite{Baum:2016oow,Rajendran:2020tmw,Budker:2021quh,Das:2022srn,Billard:2022cqd,McKeen:2022poo,McKeen:2023ztq}.

In this paper, we explore the possibility of using bilayer graphene (BLG) as the target material to detect light DM. Graphene is a two-dimensional one-atom thick hexagonal lattice of carbon atoms with unique electronic properties\,\cite{2009RvMP...81..109C}. In monolayer graphene, the electrons have a linear dispersion relation in the low energy regime. The valence and conduction bands of an intrinsic monolayer graphene touch each other at a single point in the Brillouin zone, giving rise to its semimetallic properties with point-like Fermi surface. Previously, monolayer graphene has been suggested as a detector material for sub-GeV mass DM search\,\cite{Wang:2015kya,Hochberg:2016ntt,Kim:2020bwm,Catena:2023awl,Catena:2023qkj}. The two-dimensional nature of monolayer graphene was used in some of these studies to achieve directional sensitivity of the DM scattering in the detector which may be helpful in isolating the DM signal from the background in a future experiment. Most of them depend on the removal of an electron from the graphene sheet, and a minimum energy of at least a few eV is needed for a $\pi$-band electron ejection. However, detecting sub-MeV DM requires sub-eV energy threshold target making monolayer graphene unsuitable for this purpose. Moreover, the electronic bands in a monolayer graphene do not have any energy gap which can be set as a threshold. Having an energy gap $E_g$ is crucial to suppress the thermal fluctuation noise in an experiment which can be done by cooling down the setup below the energy gap ($k_BT<E_g$). Therefore, achieving a sub-eV threshold for sub-MeV DM requires us to venture beyond monolayer graphene. 


Bernal stacked (AB stacked) BLG consists of two coupled monolayer graphene sheets, where half of the carbon atoms in the first layer are positioned directly above the hexagonal center of the second layer, while the remaining carbon atoms align directly above the carbon atoms in the second layer. The interlayer coupling between the electrons from the two layers gives BLG distinct electronic properties that are absent in monolayer graphene\,\cite{2008JPhCS.129a2002C,2013RPPh...76e6503M,2016PhR...648....1R}. 
Most importantly, a \emph{tunable} energy gap can be opened up to $\sim300\meV$ between the valence and conduction bands by externally applying a gate voltage across the BLG surface\,\cite{2009RvMP...81..109C, McCann2006:prb, Min2007:prb, Oostinga:2008natmat, Castro2007:prl}. Then electronic excitation between the bands is possible for DM scattering with sufficient energy deposition. We compute the dielectric function of BLG from the model Hamiltonian in the low-energy continuum limit, and use it to compute the DM scattering and absorption rates for two example values of sub-eV energy gap: $50$ and $100\meV$\,\cite{Knapen:2020aky,Hochberg:2021pkt,Knapen:2021bwg}. The excited electrons can be detected using an external circuit. We show that BLG can have good sensitivity for DM-electron scattering cross section and for dark photon DM mixing parameter, even with a relatively small exposure. Moreover, the intrinsic 2D nature of BLG makes it a suitable material to measure the daily modulation of DM signal expected due to Earth's rotation about its axis. This modulation can also potentially help us reject the background sources that do not share the same time variation. Finally, we provide a future detector design, and discuss possible sources of background noise.


This paper is organized as follows. In section\,\ref{sec:blg}, we give an introduction to the electronic properties of BLG and the effects of applying a gate voltage. Then in section\,\ref{sec:DM-e}, we compute its dielectric function in the isotropic limit and compute the sensitivity of a BLG detector for DM-electron scattering cross section and vector DM absorption. We show sensitivity projections and discuss the results in section\,\ref{sec:results}. Finally, we describe a possible detector design and discuss background noise in section\,\ref{sec:detector}, and conclude in section\,\ref{sec:conclusion}. We use the natural units $\hbar=c=\epsilon_0=1$ throughout this paper.

\section{Electronic properties of bilayer graphene}\label{sec:blg}
\begin{figure*}[t]
    \centering
    \includegraphics[width=1.6\columnwidth]{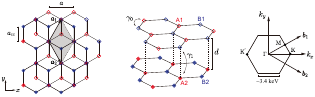}
    \caption{(\emph{Left}) Crystal structure of bilayer graphene. The unit cell is highlighted in gray and contains four sublattices: A1, B1, A2, and B2. The numbers 1 and 2 represent the layer indices for the top (empty circles) and bottom (filled circles) layers, respectively. (\emph{Center}) A schematic illustration of the stacking sequence for bilayer graphene with the hoppings represented by  $\gamma_0$ and $\gamma_1$. The B sites of the top layer (B1) are aligned with the A sites of the bottom layer (A2). (\emph{Right}) The reciprocal lattice with the reciprocal lattice vectors, and high-symmetry points $\mathrm{\Gamma}, \mathrm{K}, \mathrm{K'}$, and $\mathrm{M}$ are indicated with black dots.}
    \label{fig:gr_intro}
\end{figure*}

BLG is a two-dimensional material consisting of two stacked layers of graphene, each being a single layer of carbon atoms arranged in a hexagonal lattice. BLG can be stacked in various ways, including AA stacked graphene\,\cite{Liu:2009prl,deAndres:2008prb}, twisted BLG\,\cite{doi:10.1073/pnas.1108174108, 2018Natur.556...43C, Cao2018Cor, Andrei2020}, and so on. Among these, we investigate a prominent type known as Bernal stacked (AB stacked) BLG\,\cite{Ohta:2006sci, 2013RPPh...76e6503M} for a dark matter detector. From now on, unless otherwise specified, referring to BLG will mean Bernal stacked (AB stacked) bilayer graphene. Figure\,\ref{fig:gr_intro} shows the structure of monolayer and bilayer graphene in both real space and reciprocal lattice space. In both cases, the lattice structure is described by the primitive lattice vectors
\begin{align}
    \boldsymbol{a}_1=a\left(\frac{1}{2}, \frac{\sqrt{3}}{2}\right), 
    \quad
    \boldsymbol{a}_2=a\left(\frac{1}{2}, -\frac{\sqrt{3}}{2}\right), 
\end{align}
where $a=2.46\ \mathrm{\AA}$ is the lattice constant. The distance between adjacent carbon atoms is $a_{\mathrm{CC}}=a/\sqrt{3}=1.42\ \mathrm{\AA}$. Each layer of monolayer graphene has two inequivalent sublattices, labeled by A (red) and B (blue) in figure\,\ref{fig:gr_intro}. Therefore, in the case of bilayer graphene, there are four sublattices in the unit cell. The corresponding reciprocal lattice vectors $\boldsymbol{b}_1$ and $\boldsymbol{b}_2$ satisfy the condition $\boldsymbol{a}_i \cdot \boldsymbol{b}_j = 2\pi \delta_{ij}$ and are given as
\begin{align}
    \boldsymbol{b}_1=\frac{2\pi}{a}\left(1, \frac{1}{\sqrt{3}}\right), \quad
    \boldsymbol{b}_2=\frac{2\pi}{a}\left(1, -\frac{1}{\sqrt{3}}\right).
\end{align}

\begin{figure*}[t]
    \centering
    \includegraphics[width=1.8\columnwidth]{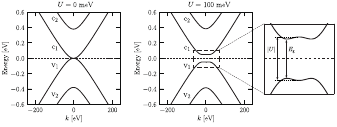}
    \caption{Band structure of bilayer graphene near the $\mathrm{K}$ or $\mathrm{K}'$ points: without any external electric field (\emph{left}) and with an external electric field (\emph{right}) inducing on-site energy differences of $U=100\meV$. The dashed line represents the Fermi energy. The band structure near the Fermi energy is magnified on the right. Here, $|U|$ indicates the energy difference between the two bands labeled by c1 and v1 at the $\mathrm{K}$ or $\mathrm{K}'$ point, and $E_g$ represents the band gap.}
    \label{fig:gr_band}
\end{figure*}

The low-energy band structure of the monolayer graphene is characterized by a two-dimensional massless Dirac equation with linear dispersion at two inequivalent hexagonal corners (often referred to as the valley) in the Brillouin zone, namely the $\mathrm{K}$ and $\mathrm{K}'$ points. For a quantitative understanding of the low-energy band structure of bilayer graphene, we introduce the continuum model Hamiltonian that describes the low-energy band structure near the valley ($\mathrm{K}$ or $\mathrm{K}'$) with the basis of four atomic sites in the unit cell (A1, B1, A2, B2):
\begin{align} \label{eq:ham_blg_con}
    H =
    \begin{pmatrix}
        \varepsilon_{\mathrm{A1}} & v\pi^{\dagger} & 0 & 0 \\
        v \pi & \varepsilon_{\mathrm{B1}} & \gamma_1 & 0 \\
        0 & \gamma_1 & \varepsilon_{\mathrm{A2}} & v\pi^{\dagger} \\
        0 & 0 & v \pi & \varepsilon_{\mathrm{B2}}
    \end{pmatrix},
\end{align}
where $\pi=\xi k_x+i k_y$ is the momentum measured from the $\mathrm{K}$ ($\xi=1$) or $\mathrm{K}'$ ($\xi=-1$) points, and $v=\sqrt{3}a\gamma_0/2~ (\approx 3 \times 10^{-3}$ for large $k$ within the continuum model) is the effective velocity, and $\gamma_0$ and $\gamma_1$ are the intralayer and interlayer coupling between the nearest neighbors, respectively, as described in figure\,\ref{fig:gr_intro}. The diagonal terms $\varepsilon_{\mathrm{A1}}$, $\varepsilon_{\mathrm{B1}}$, $\varepsilon_{\mathrm{A2}}$, $\varepsilon_{\mathrm{B2}}$ are the on-site energies of the four atomic sites in the unit cell.
Bilayer graphene consists of two stacked monolayer graphene. Thus, at the $\mathrm{K}$ and $\mathrm{K}'$ points, there exist two copies of the Dirac Hamiltonian, originating from the top and bottom monolayer graphene sheets, which correspond to the upper-left and lower-right 2$\times$2 blocks of $H$ in eq.\,(\ref{eq:ham_blg_con}). Due to the interlayer coupling $\gamma_1$, these two Dirac Hamiltonians are coupled. This results in a pair of bands with quadratic dispersion at the Fermi energy and another pair of bands that is split away of the order of $\gamma_1$ from the Fermi energy, as shown in figure\,\ref{fig:gr_band}. For the hopping parameters, we used the values $\gamma_0=3.16\eV$ and $\gamma_1=0.381\eV$\,\cite{2013RPPh...76e6503M}. More distant hoppings have a minor effect on the electronic structure, and thus exerting only a minor influence on the physics of interest in this study. Hence, we only consider the two dominant hoppings, for simplicity. As $k$ moves away from the $\mathrm{K}$ ($\mathrm{K}'$) point ($k\gtrsim100\eV$), the impact of $\gamma_1$ on the band structure diminishes. The bands start to show a linear dispersion with a band velocity $v$, similar to the monolayer case. We emphasize that the eigenenergies of the continuum model Hamiltonian in eq.\,(\ref{eq:ham_blg_con}) are isotropic with respect to the direction of the momentum $\boldsymbol{k}$. It is also important to note that the continuum model is valid roughly within the region where $k\lesssim1\keV$. Beyond this limit, the system becomes anisotropic, and a tight-binding model, which is valid across the entire Brillouin zone, is required to describe the accurate behavior of the band structure, rather than a continuum model\,\cite{McCann2013:rpp, Jung:2014prb}.

Intrinsic bilayer graphene does not have an energy gap like monolayer graphene. However, an external potential applied to bilayer graphene induces a gap\,\cite{2009RvMP...81..109C, McCann2006:prb, Min2007:prb}. Experimental studies have shown that, through doping or electric gating, a gap of up to approximately $300\meV$ can be achieved\,\cite{Oostinga:2008natmat, Castro2007:prl}. This effect can be directly modeled from the Hamiltonian in eq.\,(\ref{eq:ham_blg_con}). In the presence of an external electric field perpendicular to the graphene ($\boldsymbol{E}_{\mathrm{ext}} \parallel \boldsymbol{z}$), the same potential is applied within a single layer, but the two layers experience different potentials. Denoting the potential difference between the two layers as $U$, the on-site energies become $\varepsilon_{\mathrm{A1}}=\varepsilon_{\mathrm{B1}}=U/2$ and $\varepsilon_{\mathrm{A2}}=\varepsilon_{\mathrm{B2}}=-U/2$. The band structures of BLG without and with an external potential are depicted in figure\,\ref{fig:gr_band}. The electric field opens a gap in the low-energy band structure of bilayer graphene, creating a band with a Mexican hat structure, where the band gap is given by $E_{g}=|U|\gamma_1/\sqrt{\gamma_1^2+U^2}$. For small values of $U$, the difference between $U$ and $E_g$ is small. For example, $E_g=49.6$ meV when $U=50$ meV, and $E_g=96.7$ meV when $U=100$ meV. Thus, we will not distinguish between the values of $U$ and $E_g$ in the rest of the paper and often use both interchangeably.


\section{Dark matter-electron scattering in bilayer graphene}\label{sec:DM-e}
In this section, we describe the methods to calculate the DM scattering and absorption rates using the energy loss function of BLG.

\subsection{Kinematics of dark matter}
In any detector, DM scattering rate per unit target mass can be written as\,\cite{Trickle:2019nya,Kahn:2021ttr},
\begin{align}
    R(t) = \frac{1}{\rho_T} \frac{\rho_{\chi}}{m_{\chi}}\int d\omega \int \frac{d^3q}{(2\pi)^3} \frac{\pi\sigma(q)}{\muce^2} g(\bq,\omega,t) S(\bq,\omega). \label{eq:scattering_rate}
\end{align}
Here, $\rho_T$ is the detector material density, $\rho_\chi$ is the local DM density that we take to be $0.4\,\mathrm{GeV cm^{-3}}$\,\cite{Read:2014qva}, $\mchi$ is the DM mass, $\sigma(q)$ is the DM-electron scattering cross section, and $\muce$ is the DM-electron reduced mass. The dynamic structure factor $S(\bq,\omega)$ characterizes the response of the detector material for a momentum $\bq$ and energy deposition $\omega$.

The scattering cross section can be split into a constant factor $\schie$ and a mediator form factor $\fmed(q)$ which is momentum-dependent, such that $\sigma(q) = \schie\fmed(q)^2$, where
\begin{equation}\label{eq:form_factor}
    \schie\equiv \dfrac{\mu_{\chi e}^2}{\pi}\left(\dfrac{g_eg_\chi}{q_0^2+\mphi^2}\right)^2\,, \quad
    \fmed(q) = \dfrac{q_\mathrm{0}^2+\mphi^2}{q^2+\mphi^2}\,.
\end{equation}
Here, $q_0=\alpha m_e$ is a reference momentum, $\mphi$ is the mediator mass, and $g_e,g_\chi$ are the couplings of the mediator with electron and DM, respectively. As mentioned in section\,\ref{sec:blg}, we will compute the BLG response function assuming the isotropic approximation which is not valid for large $q$ and $\omega$. Hence in this paper, we will always assume a massless mediator, i.e. $q\gg\mphi$.

The function $g(\bq,\omega,t)$ is the only time-varying quantity in eq.(\ref{eq:scattering_rate}) and contains the directional information of the scattering rate. It is obtained by integrating the DM velocity distribution, which is assumed to be Maxwellian with a mean velocity $v_0=230\kms$ and cut off at the local galactic escape velocity $v_{\rm esc}$. A minimum velocity $v_-$ is taken for a certain $q$ and $\omega$\,\cite{Trickle:2019nya,Kahn:2021ttr},
\begin{align}\label{eq:g_func}
    \nonumber & g(\bq,\omega,t) = \dfrac{2\pi^2 v_0^2}{qN_0}
    \left[\exp\left(-\dfrac{v_-(\bq,t)^2}{v_0^2}\right) - \exp\left(-\dfrac{v_{\rm esc}^2}{v_0^2}\right)\right]
    \,,\\[1ex]
    &v_-(\bq,t) = \text{min}\{v_\mathrm{esc}, \frac{\omega}{q} + \frac{q}{2\mchi} + \hat{q}\cdot\bv_\mathrm{lab}(t)\}\,.
\end{align}
Here $N_0$ is the normalization constant of the truncated Maxwell distribution 
\begin{align}
    N_0 = \pi^{3/2}v_0^3\left[\mathrm{erf}\left(\frac{v_\mathrm{esc}}{v_0}\right) - \frac{2v_\mathrm{esc}}{\pi^{1/2}v_0}\exp\left(-\frac{v_\mathrm{esc}^2}{v_0^2}\right)\right]\,,
\end{align}
and $\bv_\mathrm{lab}(t)$ is the time-dependent velocity of the laboratory on the Earth in the galactic frame,
\begin{equation}\label{eq:v_lab}
    \bv_\mathrm{lab}(t) = |\bv_E|\begin{pmatrix}
        \sin\theta_e \sin\phi\\
        \sin\theta_e \cos\theta_e (\cos\phi-1)\\
        \cos^2\theta_e + \sin^2\theta_e \cos\phi
    \end{pmatrix}\,,
\end{equation}
with $\phi=2\pi\times (t/24\,\mathrm{h}), \theta_e=42^\circ$, and $|\bv_E|=240\kms$.  
All directional information of the DM flux is included in $v_-(\bq,t)$.

The dynamic structure factor $S(\bq,\omega)$ can be computed in terms of the dielectric function $\varepsilon(\bq,\omega)$ using the fluctuation-dissipation theorem as long as the DM-electron interaction is weak\,\cite{Knapen:2020aky,Hochberg:2021pkt,Knapen:2021bwg}. We describe this method in the next section.

\subsection{Calculation of the dynamic structure factor}
Equation (\ref{eq:scattering_rate}) above needs to be modified appropriately as we are going to use BLG in this study, which is a 2D material. 
The dynamic structure factor encodes the information of the correlation of the density fluctuations in a system at a momentum $\bq$ and energy $\omega$. It can be expressed using the electron density operator as the Fourier transform of the density-density correlation function\,\cite{giuliani:2005quantum, girvin:2019modern}, i.e.
\begin{align}\label{eq:dsf}
    S(\boldsymbol{q},\omega) = \frac{1}{V} \int dt e^{i\omega t} \langle\hat{n}(\boldsymbol{q},t) \hat{n}(-\boldsymbol{q},0)\rangle\,. 
\end{align}
Here, $\hat{n}(\boldsymbol{q},t)$ is the electron density operator at a momentum $\bq$ and time $t$:
\begin{align}\label{eq:denstiy_ft}
    \hat{n}(\boldsymbol{q},t) = \int d^3 r\, e^{-i \boldsymbol{q} \cdot \boldsymbol{r}} \hat{n}(\boldsymbol{r},t)\, ,
\end{align}
where $\hat{n}(\boldsymbol{r},t)$ is the electron density operator in real space. For 2D materials like BLG, the electron density is confined to the 2D space (near $r_z=0$). We can assume that $\hat{n}(\boldsymbol{r},t)$ can be separated into the material's in-plane ($r_x$, $r_y$) and out-of-plane ($r_z$) components. Then the electron density operator can be written as
\begin{align}
    \hat{n}(\boldsymbol{r}, t) = \hat{n}_{\mathrm{2D}}(\boldsymbol{r}_{\parallel}, t) \left[ \frac{1}{d}\Theta(d/2-|r_z|)\right]. \label{eq:density_sep}
\end{align}
Here, we make the assumption that the electron density remains constant between the two layers of BLG, defined within the range $-d/2 < r_z < d/2$, where $d = 3.35$\,$\mathrm{\AA}$ represents the interlayer distance. The spatial coordinates in the in-plane direction are denoted as $\boldsymbol{r}_{\parallel} = (r_x, r_y)$, and $\Theta(x)$ represents the Heaviside step function.

Note that we have assumed a step function-like profile for the real-space electron density, as such an assumption does not significantly impact the results. Given that the DM mass is on the sub-MeV regime, the order of magnitude for $q$ is roughly in the sub-keV range, implying $qd \ll 1$. This implies that the length scale of the interlayer separation in bilayer graphene is much smaller than the inverse of the typical momentum transfer $\sim1/q$, rendering the detailed structure of the real-space density profile less crucial for our considerations.
Plugging eq.\,(\ref{eq:density_sep}) into eq.\,(\ref{eq:denstiy_ft}), we get
\begin{align}
    \hat{n}(\boldsymbol{q}, t) =& \int d^3r e^{-i\boldsymbol{q}\cdot\boldsymbol{r}} \hat{n}(\boldsymbol{r}, t) \nonumber \\[1ex]
    =& \left[\int d^2r_{\parallel} e^{-i\boldsymbol{q}_{\parallel}\cdot\boldsymbol{r}_{\parallel}} n_{\mathrm{2D}}(\boldsymbol{r}_{\parallel}, t) \right]\nonumber\\[1ex]
    & \times\left[\int dr_z e^{-i q_z r_z} \frac{1}{d}\Theta(d/2-|r_z|) \right] \nonumber \\[1ex]
    =& \hat{n}_{\mathrm{2D}}(\boldsymbol{q}_\parallel, t) \left(\frac{\sin{q_zd/2}}{q_zd/2} \right).
\end{align}
Using this, we can rewrite eq.\,(\ref{eq:dsf}) in terms of the electron density operator in 2D as
\begin{align}
    S(\boldsymbol{q},\omega) =& \frac{1}{V} \int dt e^{i\omega t} \langle\hat{n}(\boldsymbol{q},t) \hat{n}(-\boldsymbol{q},0)\rangle \nonumber \\[1ex]
    =& \frac{1}{d} \left( \frac{1}{A} \int dt e^{i\omega t} \langle n_{\mathrm{2D}}(\boldsymbol{q}_{\parallel}, t) n_{\mathrm{2D}}(-\boldsymbol{q}_{\parallel}, 0)\rangle \right)\nonumber\\[1ex] &\times\left(\frac{\sin{q_zd/2}}{q_zd/2} \right)^2 \nonumber \\[1ex]
    =& \frac{1}{d} S_{\mathrm{2D}}(\boldsymbol{q}_{\parallel},\omega) \left(\frac{\sin{q_zd/2}}{q_zd/2} \right)^2,
\end{align}
where $V=Ad$ is the 3D volume, with $A$ being the 2D volume or surface area. Here, $S_{\mathrm{2D}}(\boldsymbol{q}_{\parallel},\omega)$ denotes the dynamic structure factor of the 2D material, and the following square of the sinc function contains the variation in the $q_z$-direction.
Combining the aforementioned information, the detector component $\frac{1}{\rho_T} S(\boldsymbol{q},\omega)$ in the DM-electron scattering rate in eq.\,(\ref{eq:scattering_rate}) transforms as
\begin{align}
    \frac{1}{\rho_T} S(\boldsymbol{q},\omega) \ \rightarrow \ \frac{1}{\sigma_T} S_{\mathrm{2D}}(\boldsymbol{q}_{\parallel},\omega) \left(\frac{\sin{q_zd/2}}{q_zd/2} \right)^2,
\end{align}
where $\sigma_T = 1.53\times10^{-7} \mathrm{g/cm^2}$ is the surface density of BLG.
From here on, unless otherwise specified, the dynamic structure factor will refer to $S_{\mathrm{2D}}$.
\begin{figure*}[t]
    \centering
    \includegraphics[width=1.95\columnwidth]{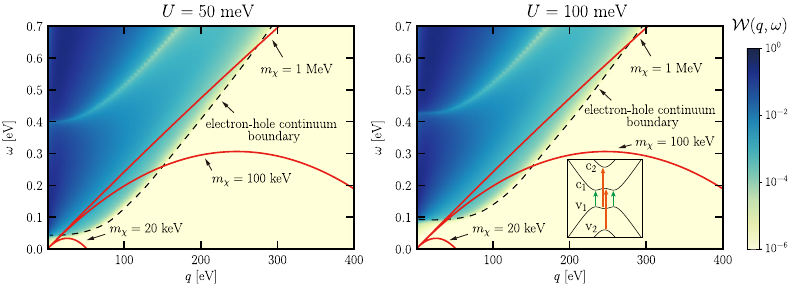}
    \caption{The color plot of the energy loss function (ELF) for BLG with $U=50\meV$ (\emph{left}) and $100\meV$ (\emph{right}), represented in the $q$ - $\omega$ plane. The black dashed line indicates the boundary of the electron-hole continuum, with the more intensely colored region above signifying the continuum itself. The kinematically-allowed boundaries for DM with mass $m_{\chi}=20, 100$, and $1000 \keV$ are depicted by the red solid lines. The inset in the right panel figure shows the different electronic transitions that give rise to the features of the ELF. See section\,\ref{sec:dielectric_blg} for more details.}
    \label{fig:gr_elf}
\end{figure*}
Specifically, the dynamic structure factor can be written as the imaginary (dissipative) part of the 2D density-density response function $\chi_{\mathrm{2D}}(\boldsymbol{q}_{\parallel}, \omega)$ derived from the fluctuation-dissipation theorem\,\cite{giuliani:2005quantum,girvin:2019modern}:
\begin{align} \label{eq:dsf-fd}
    S_{\mathrm{2D}}(\boldsymbol{q}_{\parallel}, \omega) = -\frac{2}{1-e^{-\beta \omega}} \mathrm{Im} \left[\chi_{\mathrm{2D}}(\boldsymbol{q}_{\parallel}, \omega)\right],
\end{align}
where $\beta=1/(k_B T)$. The 2D density-density response function in the frequency-momentum space is the Fourier transform of the 2D retarded density correlation function given by
\begin{align} \label{eq:density-correlation}
    \chi_{\mathrm{2D}}(\boldsymbol{q}_{\parallel}, \omega) = -\frac{i}{A} \int dt\, \Theta(t) e^{i\omega t} \left< \left[ \hat{n}_{\mathrm{2D}}(\boldsymbol{q}_{\parallel},t), \hat{n}_{\mathrm{2D}}(-\boldsymbol{q}_{\parallel},0) \right]\right>.
\end{align}
Note that the expectation value $\langle \cdots \rangle$ refers to the thermal average over many-body states, taking into account the Coulomb interaction between the electrons. It is often practical to introduce the dielectric function of the detector. The dielectric function, in the frequency-momentum space, is defined as the ratio of the screened potential to the bare potential in a material. The 2D version of the dielectric function is given by
\begin{align} \label{eq:dielec-rpa}
    \frac{1}{\varepsilon(\boldsymbol{q}_{\parallel}, \omega)} = 1 + v_{\mathrm{C}}(\boldsymbol{q}_{\parallel})\chi_{\mathrm{2D}}(\boldsymbol{q}_{\parallel}, \omega)\, ,
\end{align}
where $v_{\mathrm{C}}(q)=\dfrac{e^2}{2\varepsilon_r q}$ is the 2D Coulomb interaction. Here, $\varepsilon_r$ is the dielectric constant of the medium which we assume to be a vacuum. At zero temperature, the dynamic structure factor in eq.\,(\ref{eq:dsf-fd}) becomes
\begin{align} \label{eq:elf_def}
    & S_{\mathrm{2D}}(\boldsymbol{q}_{\parallel}, \omega) = -2 \Theta (\omega) \mathrm{Im} \chi_{\mathrm{2D}}(\boldsymbol{q}_{\parallel},\omega) = \frac{2 \Theta(\omega)}{v_{\mathrm{C}}(\boldsymbol{q}_{\parallel})} \mathcal{W}(\boldsymbol{q}_{\parallel}, \omega),\nonumber\\[1ex]
    &\mathcal{W}(\boldsymbol{q}_{\parallel}, \omega) = \mathrm{Im}\left[-\frac{1}{\varepsilon(\boldsymbol{q}_{\parallel}, \omega)} \right].
\end{align}
This implies a direct connection between the dynamic structure factor and the imaginary part of the inverse dielectric function, termed the energy loss function (ELF) $\mathcal{W}(\boldsymbol{q}, \omega)$. The ELF represents electronic energy loss in materials and is thus related to an experimentally measurable quantity. The information about the response to a small density perturbation is contained in the dielectric function.

To calculate the ELF, we can use the random phase approximation (RPA) to compute the expectation value in eq.\,(\ref{eq:density-correlation}). Diagrammatically, the RPA corresponds to an infinite series of the non-interacting polarization function connected by the Coulomb interaction lines, neglecting other higher-order terms\,\cite{mahan:2000many, bruus:2004many, giuliani:2005quantum, girvin:2019modern}:
\begin{align} \label{eq:pol-rpa}
    &\chi_{\mathrm{2D}}(\boldsymbol{q}_{\parallel}, \omega) = \frac{\chi_{\mathrm{2D}}^{(0)}(\boldsymbol{q}_{\parallel}, \omega)}{\varepsilon(\boldsymbol{q}_{\parallel}, \omega)}\,,\nonumber\\[1ex]
    &\varepsilon(\boldsymbol{q}_{\parallel}, \omega) = 1-v_{\mathrm{C}}(\boldsymbol{q}_{\parallel}) \chi_{\mathrm{2D}}^{(0)}(\boldsymbol{q}_{\parallel}, \omega)\,,
\end{align}
where the response function in eq.\,(\ref{eq:density-correlation}) and the dielectric function in eq.\,(\ref{eq:dielec-rpa}) are expressed in terms of the non-interacting polarization function $\chi^{(0)}_{\mathrm{2D}}(\bq, \omega)$  using the RPA. This approach is known to be valid in the weak coupling (typically high density) limit or in the large fermion flavor limit.
$\chi^{(0)}_{\mathrm{2D}}(\bq, \omega)$ is given by\,\cite{giuliani:2005quantum}
\begin{align} \label{eq:pol-spectral}
    \chi^{(0)}_{\mathrm{2D}}(\boldsymbol{q}_{\parallel}, \omega) = g \sum_{n,m} \int \frac{d^2k}{(2\pi)^2} &\frac{f_{\boldsymbol{k},n} - f_{\boldsymbol{k}+\boldsymbol{q}_{\parallel},m}}{\omega +\varepsilon_{\boldsymbol{k},n} - \varepsilon_{\boldsymbol{k}+\boldsymbol{q}_{\parallel},m} + i\eta}\nonumber\\[1ex]
    &\times |\langle \boldsymbol{k},n|\boldsymbol{k}+\boldsymbol{q}_{\parallel},m \rangle|^2,
\end{align}
where $\varepsilon_{\boldsymbol{k},n}$ and $f_{\boldsymbol{k},n}=1/[1+e^{\beta(\varepsilon_{\boldsymbol{k},n}-\mu)}]$ are the eigenenergy and the Fermi-Dirac distribution for the wave vector $\boldsymbol{k}$ and band index $n$, respectively, $g=4$ is the spin-valley degeneracy factor for graphene, and $\mu$ is the chemical potential, $\eta$ is a phenomenological broadening parameter proportional to the inverse lifetime (or decay width) of quasiparticles which takes $\eta\rightarrow0^+$ for the clean limit. For numerical calculations, we set $\eta=1$ meV, which is the typical order of magnitude for BLG\,\cite{PhysRevLett.127.117701,PhysRevLett.102.176804,PhysRevLett.87.267402}. We note that variations in $\eta$ have a negligible impact on the overall results.

In this work, we will also consider DM absorption in BLG. To this end, we will show the results for a dark photon model where the DM mixes with the Standard Model (SM) photon through a small mixing parameter $\kappa$,
\begin{align}
    \mathcal{L}\supset \kappa F_{\mu\nu}F'^{\mu\nu}\,,
\end{align}
where $F^{\mu\nu}$ and $F'^{\mu\nu}$ are the field strengths of the SM photon and the dark photon, respectively. Because of the mixing between dark photon DM and the Standard Model photon, DM absorption rate is again related to the ELF of the material rescaled by the mixing parameter $\kappa$ as follows\,\cite{Mitridate:2021ctr},
\begin{align}\label{eq:absorption}
    R_{\rm abs} = \kappa^2\rho_\chi\dfrac{d}{\sigma_T}\,\mathrm{Im}\left[-\frac{1}{\varepsilon(q, \omega)} \right]\,,
\end{align}
with $\omega=\mchi$ and $q=\mchi v$. Because $q\ll\omega$ for DM absorption, it is more convenient to calculate the dielectric function in 2D using the following expression\,\cite{jackson, bruus:2004many}
\begin{align}
    \varepsilon(q\rightarrow0, \omega) = 1+\dfrac{i}{2\varepsilon_r} \dfrac{q}{\omega} \sigma(\omega).
\end{align}
Here $\varepsilon_r$ denotes the dielectric constant of the medium which we assume to be vacuum, and the optical conductivity $\sigma(\omega)$ of BLG is calculated using the Kubo formula\,\cite{mahan:2000many}.
We will show projection of constraint on the mixing parameter $\kappa$.






\section{Results}\label{sec:results}
We now have all the tools needed to calculate the ELF of BLG, and the DM scattering and absorption rates. In this section, we will show and explain our main results.
\subsection{Dielectric properties of bilayer graphene}\label{sec:dielectric_blg}

The ELF of BLG in the presence of a potential, calculated through eqs.\,(\ref{eq:elf_def})-(\ref{eq:pol-spectral}), is shown as a color plot in the $q$ - $\omega$ plane in figure\,\ref{fig:gr_elf}. Since the low-energy band structure described by the continuum model Hamiltonian in eq.\,(\ref{eq:ham_blg_con}) is isotropic, the ELF depends only on the magnitude of $\bq$ and not its direction.
\begin{figure*}[t]
    \centering
    \includegraphics[width=\columnwidth]{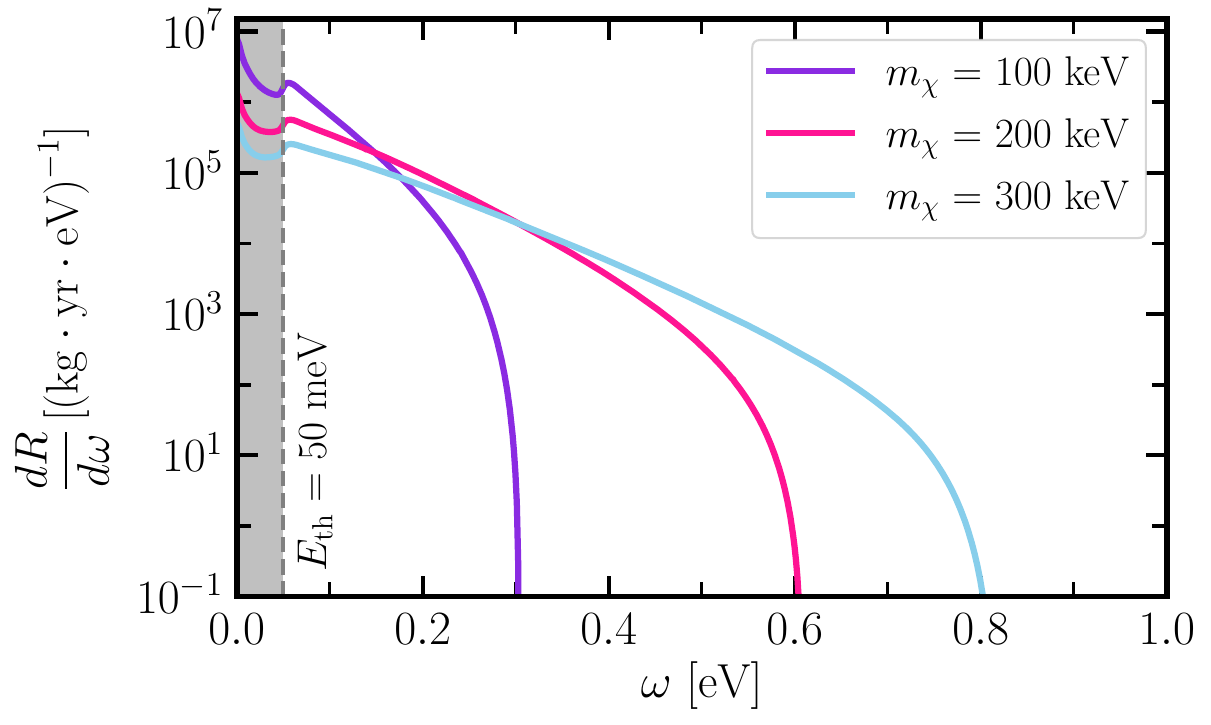}~\includegraphics[width=\columnwidth]{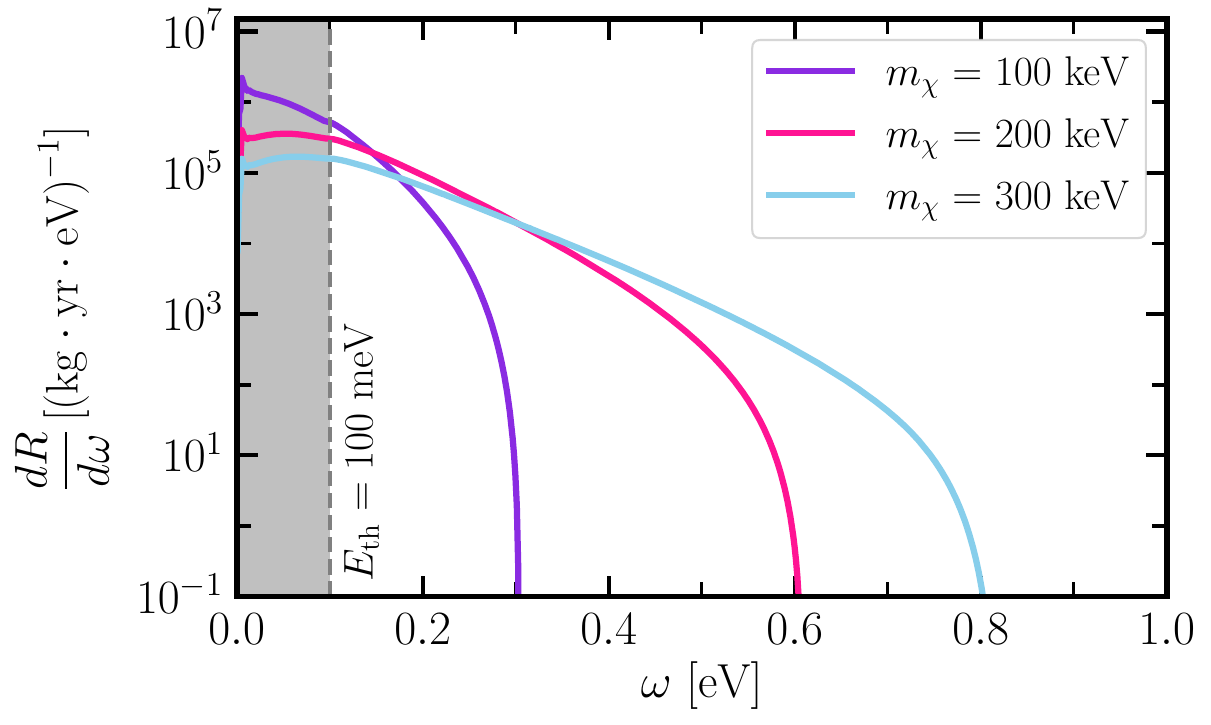}
    \caption{Energy spectrum of DM scattering rate in bilayer graphene with two band gaps $U=50\meV$ (\emph{left}) and $100\meV$ (\emph{right}) for $\mchi=100$ (violet), 200 (pink), and $300\keV$ (sky blue). A massless mediator and $\schie=10^{-38}\cmsq$ are assumed in all cases. The energy thresholds are marked with a vertical gray line. The gray-shaded below-threshold regions are not taken into account in our results.}
    \label{fig:rate_spectrum}
\end{figure*}

The magnitude of the ELF is related to the electronic energy loss at the momentum transfer $q$ and energy transfer $\omega$. Especially when there is a gap and the Fermi energy lies within it, the origin of the ELF is the interband transitions of electrons. This represents cases where momentum transfer by $q$ and energy transfer by $\omega$ are possible from an occupied band to an empty band, and this region is referred to as the electron-hole continuum\,\cite{giuliani:2005quantum}. For example, in the region where $\omega < E_{g}$, the ELF is close to zero, which means the absence of available states for excitation below the gap. The boundary of the electron-hole continuum is indicated by a dashed line in figure\,\ref{fig:gr_elf}. Additionally, for $q\gtrsim100\keV$, the boundary of the electron-hole continuum in $q$ - $\omega$ plane shows a linear dispersion with the velocity $v$ in eq.\,(\ref{eq:ham_blg_con}). We note that since the Fermi energy lies within the gap, implying no states exist at that energy, contributions from intraband transitions or plasmons to the ELF are absent in this case.

The red curves indicate the kinematic phase space of DM of mass $m_{\chi}$. Here, the mean DM velocity is assumed to be $v_0=230\kms$. In the phase space where there is an overlap between this kinematically-allowed region and the dark areas of the BLG's ELF, DM-electron scattering occurs. If there is no overlap, the scattering rate vanishes. For example, in the case where $m_{\chi}=20\keV$, the kinematically-allowed region exists outside the electron-hole continuum, leading to no DM-electron scattering. This means that the $U$ applied to the BLG determines the minimum detectable mass threshold for DM. The continuum model we employed is valid for ranges approximately $q\lesssim1\keV$ and $\omega\lesssim2\eV$. For DM with a mass $m_{\chi}=1\MeV$, the kinematically-allowed boundary extends into regions with $q>1\keV$. However, the region contributing to the DM-electron scattering rate, that is, the area overlapping with the electron-hole continuum, is approximately $q\lesssim0.6\keV$. Therefore, the description using the continuum model Hamiltonian remains valid. Consequently, our results are relevant for DM mass range up to approximately $m_{\chi}\lesssim1\MeV$.

For larger values of $m_{\chi}$, a tight-binding model that is valid across the entire Brillouin zone is required, rather than a continuum model. Then the band structure no longer remains isotropic in the momentum space, leading to the loss function developing a directional dependence on $\bq$. Additionally, when the magnitude of the $q$ exceeds the size of the Brillouin zone, considerations like the local field effects\,\cite{giuliani:2005quantum,Knapen:2021prd} become necessary in the determination of the dielectric function. On the low mass end, we are limited by the finite value of $U$. Energy deposition below $U$ is not possible which will help reduce low energy thermal fluctuations.
\begin{figure*}[t]
    \centering
    \includegraphics[width=1.03\columnwidth]{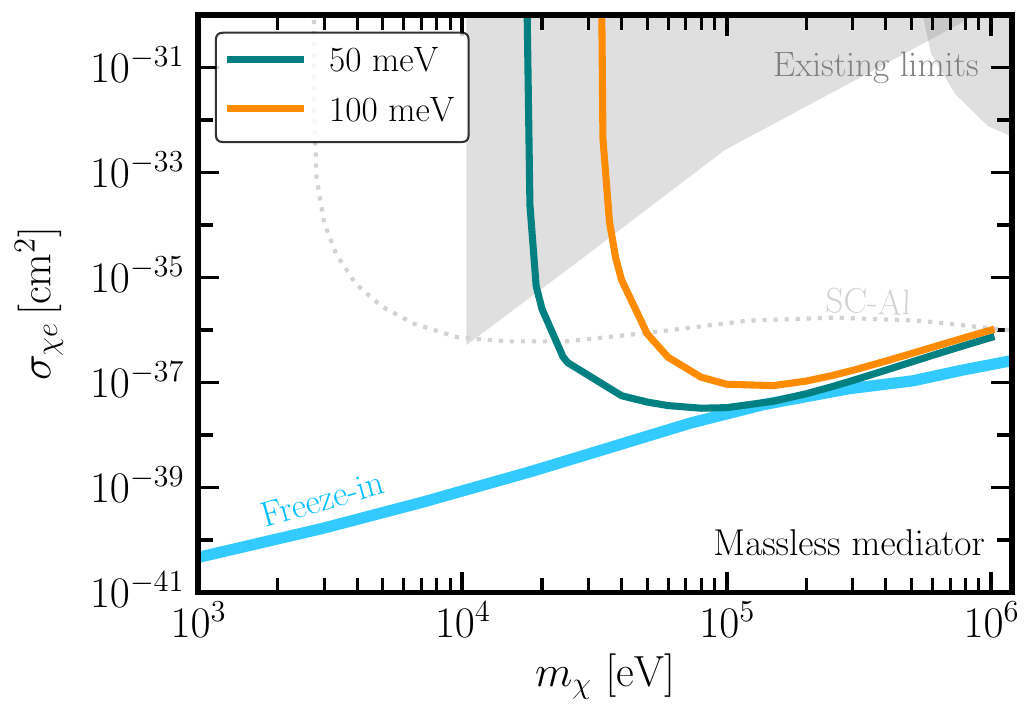}~\includegraphics[width=\columnwidth]{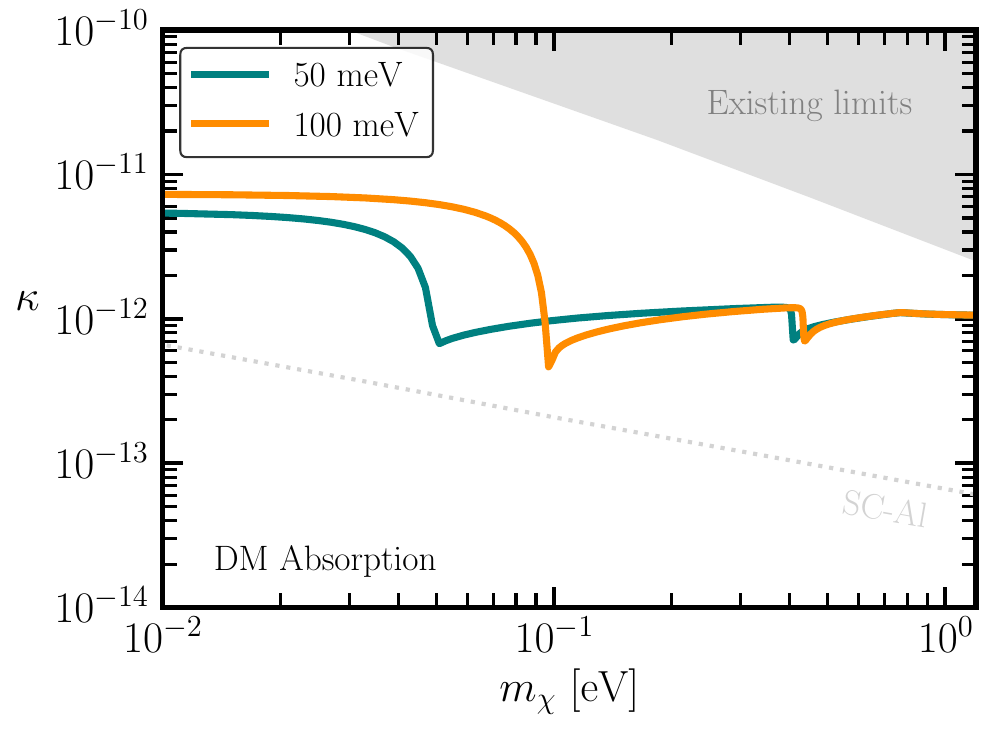}
    \caption{Projected sensitivities on DM-electron scattering cross section $\schie$ (\emph{left}) and DM mixing parameter $\kappa$ for absorption (\emph{right}) in bilayer graphene with two energy thresholds $E_\mathrm{th}=50$ and $100\meV$ assuming 3 events in 10 mg-yr exposure with no background. We use scattering rate averaged over a 24 hour period. The blue band marked with `Freeze-in' in the left panel is the region of theoretical interest. For comparison, we also show the projection for superconducting aluminum (SC-Al) as a gray, dotted line\,\cite{Hochberg:2021yud}. Bilayer graphene is superior to superconducting aluminum for DM scattering in the $20\keV-\MeV$ mass range. The dips appearing for the DM absorption sensitivity curve in the right panel are due to opening up of different band transitions of electrons. The gray-shaded regions are excluded by other laboratory and astrophysical constraints\,\cite{SENSEI:2020dpa,DAMIC:2019dcn,SuperCDMS:2020aus,FUNKExperiment:2020ofv,XENON:2019gfn,An:2014twa,An:2020bxd,Das:2021lcr,Nguyen:2021cnb,Buen-Abad:2021mvc}.}
    \label{fig:blg_limit}
\end{figure*}

Having discussed the properties of the ELF of bilayer graphene, we now compute the DM scattering rate following eq.\,(\ref{eq:scattering_rate}). We show the scattering rate as a function of energy in figure\,\ref{fig:rate_spectrum} for $\mchi= 100, 200,$ and $300\keV$ for two values of $U=50, 100\meV$ and cross section $\schie=10^{-38}\cmsq$. These thresholds are marked with a vertical gray line. As mentioned before, we will only show results for the massless mediator case. From figure\,\ref{fig:rate_spectrum}, we can see that the rates are maximum near the thresholds and decreases toward higher energies. This maximum rate comes from the electronic transitions from the valence to the conduction bands $v_1\to c_1$ as shown in the inset in the right panel of figure\,\ref{fig:gr_elf} (green arrows). Moreover, the overall rate increases for heavier DM. This is understandable from the overlap between the DM kinematic phase space and the ELF of BLG in the $q$ - $\omega$ plane in figure\,\ref{fig:gr_elf}. For low mass DM, the overlap region is small and limited toward the low energy corner of the plane, hence the scattering rate is maximum around $E_g$. For heavier mass, the overlap region is extended to higher energies and has support from the additional $v_1\to c_2$ and $v_2\to c_1$ electronic transitions marked with the orange arrows in the inset of figure\,\ref{fig:gr_elf}. This broadens the spectrum toward higher energy for heavier mass DM as can be seen in figure\,\ref{fig:rate_spectrum}. Finally, we note that the rates are higher for the case $E_g=50\meV$ because the ELF in this case peaks at a relatively lower energy ($v_1\to c_1$ transition) yielding a greater overlap with the DM kinematic phase space. This is also evident from figure\,\ref{fig:gr_elf}, indicating that the sensitivity could further improve for  smaller $E_g$. However, a smaller value of $E_g$ will also be subject to the technical challenge of maintaining the same gate voltage over a large volume of the target. The optimal value of $E_g$ will thus depend on the details of the experimental design.

Before ending this section, we want to comment on the additional below-threshold peaks that appear in the gray-shaded region in figure\,\ref{fig:rate_spectrum}. These are due to the small but nonzero value of the ELF in the $\omega<E_g$ energy regime which, in turn, is a result of our assumption of a nonzero decay width $\eta$ of the excited electron-hole pairs in eq.\,(\ref{eq:pol-spectral}). This small value of the ELF, combined with the diverging nature of the form factor $\fmed(q)\propto 1/q^2$ at small $q$, gives rise to these additional peaks. However, we note that if $\eta$ were zero, the ELF would completely vanish below $\omega<E_g$, resulting in a zero value of $dR/d\omega$ in this energy range. In our calculation, we do not include the energy range below $\omega<E_g=E_\mathrm{th}$, and assume the thresholds to be the same as the band gap in each case.

\subsection{Projected sensitivity \& daily modulation}\label{sec:sensitivity}
In this section, we will explore the future projections with bilayer graphene as the detector to search for DM. We envisage an experimental setup similar to the ones developed for photon detectors with monolayer and bilayer graphene\,\cite{2013NatCo...4.1987B,2013NatPh...9..248T,2014NatNa...9..780K,s16091351,DeSanctis:2018tkb,doi:10.1021/acs.nanolett.0c00373}. In Sec.\,\ref{sec:detector}, we will outline a detector design. To estimate the future sensitivity in this section, we assume an exposure of 10 mg-yr.

\begin{figure*}[t]
    \centering
    \includegraphics[width=\columnwidth]{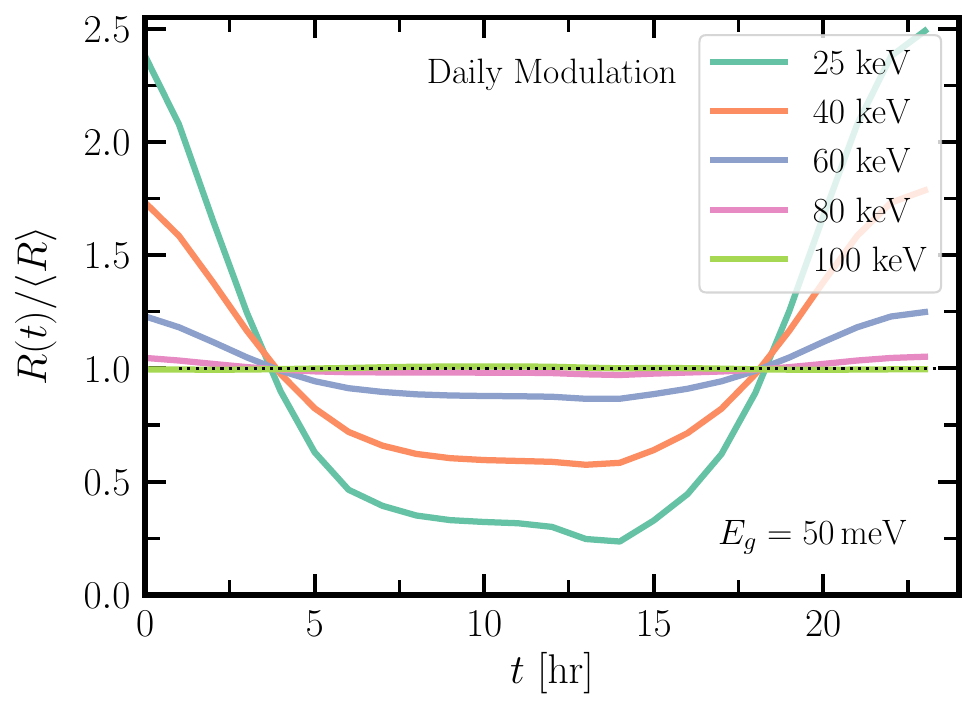}~\includegraphics[width=\columnwidth]{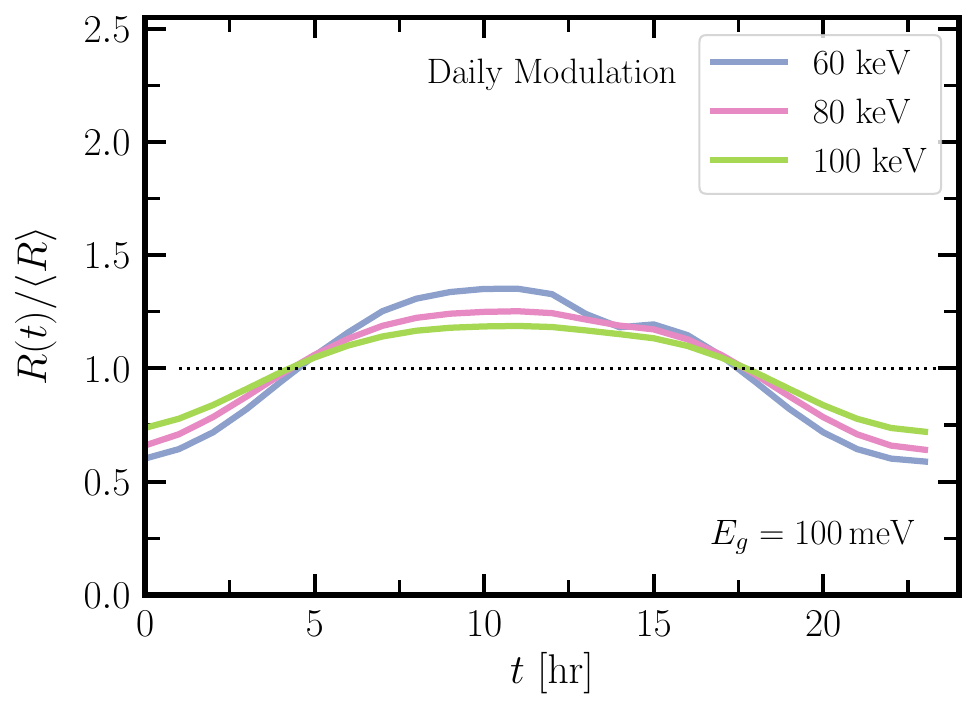}
    \caption{Daily modulation of the scattering rate in bilayer graphene for a few DM masses with $E_g=50$ (\emph{left}) and $100\meV$ (\emph{right}). The modulation is more pronounced for lighter DM as the typical energy transfer is closer to the threshold, and it is milder for lower threshold $50\meV$ compared to $100\meV$ for a fixed DM mass due to the same reason.}
    \label{fig:daily_mod}
\end{figure*}

We show the projected sensitivity for DM-electron scattering in the left panel for the case of a massless mediator, and DM absorption in the right panel of figure\,\ref{fig:blg_limit}. As mentioned before, we show the results for two values of $U=50\meV$ (orange) and $100\meV$ (teal) assuming a cutoff of 3 events for an exposure of 10 mg-yr for both cases with no background. We lose sensitivity below $\mchi\simeq 20$ and $35\keV$, respectively, for the two band gaps which can be understood from their respective energy thresholds. For comparison, we also show the projection for superconducting aluminum as a gray, dotted line\,\cite{Hochberg:2021yud}. The left panel of figure\,\ref{fig:blg_limit} shows that BLG performs better than superconducting aluminum for DM scattering in the $10\keV-\MeV$ mass range for the same amount of exposure. With the assumed 3 events per 10 mg-yr exposure limit, BLG sensitivity can reach $\schie\simeq10^{-38}\cmsq$ for $\mchi\simeq60\keV$, and hence, the band of benchmark freeze-in model marked in blue in figure\,\ref{fig:blg_limit}. We limit our calculation to below $\mchi=1\MeV$ because of the use of the continuum limit, and also because electron ionization may become important for DM masses above $1\MeV$.

In the case of DM absorption, we show our sensitivity projections on the mixing parameter $\kappa$ of dark photon DM in the mass range $0.01<\mchi<1\eV$ as the whole DM mass energy is absorbed ($\omega=\mchi$). It is computed using eq.\,(\ref{eq:absorption}) and the same limit 3 events per 10 mg-yr exposure with no background. We see that the sensitivity can reach $\kappa\simeq10^{-12}$ as shown in the right panel of figure\,\ref{fig:blg_limit}. The dips in the DM absorption sensitivity curve are coming from the opening up of electronic transitions between different bands, such as $v_1\to c_1$, $v_1\to c_2$, or $v_2\to c_1$, in the ELF as was discussed in section\,\ref{sec:dielectric_blg}. The gray-shaded regions are excluded by existing laboratory and astrophysical constraints\,\cite{SENSEI:2020dpa,DAMIC:2019dcn,SuperCDMS:2020aus,FUNKExperiment:2020ofv,XENON:2019gfn,An:2014twa,An:2020bxd}. We computed the sensitivity of BLG for pseudoscalar DM absorption too which is also related to the conductivity\,\cite{Mitridate:2021ctr}. However, we do not show it here as it does not reach a coupling strength that is not already constrained by other astrophysical observations.



Another important feature of BLG detector is the daily modulation of the DM scattering rate. This arises from the change in the direction of the incoming DM wind due to the rotation of the Earth about its axis. The two-dimensional nature of BLG facilitates the daily modulation even more as it makes the detector response $S(\bq,\omega)$ inherently anisotropic. The time modulation information in the scattering rate is included in the $g(\bq,\omega,t)$ factor in eq.(\ref{eq:scattering_rate}). In figure\,\ref{fig:daily_mod}, we show the daily modulation of the scattering rate normalized by the average, $R(t)/\langle R\rangle$ over a 24 hr period for a few DM masses in BLG with $E_g=50, 100\meV$. The rate varies by more than double for $\mchi\lesssim 30\keV$ and by smaller amount for heavier mass in the $50\meV$ case. The amount of modulation is more for lighter DM mass. This is understandable as the typical energy transfer is closer to the threshold for lighter DM and a small change in the relative velocity of DM can affect the scattering rate significantly. Such daily modulation of DM scattering rate can be leveraged to isolate the DM signal in a future experiment from various background scattering events from cosmic rays, radioactivity, or any internal systematic noise that do not share such time variation. The difference in the phase of the modulation between the left and right panels of figure\,\ref{fig:daily_mod} is due to the different DM phase space-BLG response function overlap regions in the $q-\omega$ plane. We note that on top of the daily modulation, there is also an annual modulation expected from the rotation of the Earth around the sun resulting in a slightly different relative velocities of the DM wind throughout the year. We do not calculate it in this work.

In passing, we want to point out that the DM in the Milky Way halo may have a kinematic substructure in the solar neighborhood with a different density and  velocity distribution. In such a case, the sensitivity projections and the daily modulation may get affected because of the substructure\,\cite{Buch:2020xyt,Herrera:2021puj,Maity:2022enp}.
\begin{figure*}[t]
    \centering
    \includegraphics[width=0.42\textwidth]{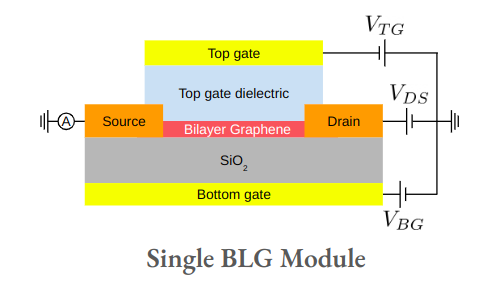}\qquad\qquad\includegraphics[width=0.49\textwidth]{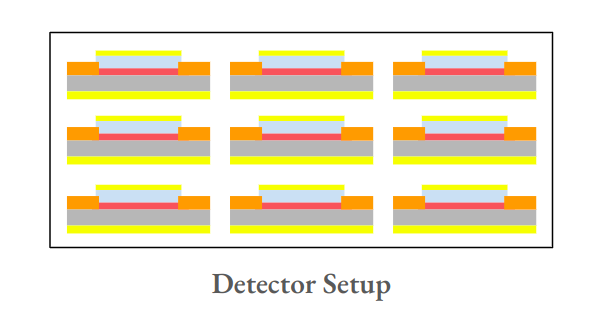}
    \caption{A schematic design of a DM detector with BLG as the target material. \textit{(Left)} A single module of the detector with the BLG (red) on an insulator substrate $\mathrm{SiO_{2}}$ (gray). The voltages $V_{TG}$, $V_{DS}$, and $V_{BG}$ are applied to the top gate, drain-source, and bottom gate, respectively, to control the electronic properties of the BLG. \textit{(Right)} The scaled up detector setup made up of an array of modules. The modular design of the detector will help scale up the exposure as needed.}
    \label{fig:detector_schematic}
\end{figure*}

\section{Experimental design}\label{sec:detector}
In this section, we will provide a design proposal of a conceptual detector using BLG, and briefly discuss the possible sources of background.

\subsection{Detector configuration}

We envisage a modular way of building a detector with BLG as the target material. A plausible way is to prepare sheets of BLG on a semiconductor substrate and connect with metallic electrodes to form a single module. Then multiple of such modules can be assembled to scale up the target mass and achieve a large exposure. A schematic design of such a module is shown in the left panel of figure\,\ref{fig:detector_schematic}. The BLG will be placed between substrates, such as $\mathrm{SiO_2}$, and connected to two metallic electrodes at the opposite ends. The gating voltage $V_{TG}$ and $V_{BG}$ can be applied at the top and bottom of the substrate as shown in figure\,\ref{fig:detector_schematic} with a metal layer covering the whole substrate to make sure a uniform electric field throughout the BLG sheet. 


We can realistically assume each BLG sheet will have a $500\,\mum\times 500\,\mum$ surface area. Note that this is still a conservative choice given that a $\sim$ cm size single crystal BLG growth technique is already available\,\cite{Nguyen2020,patent_Jeong}. A target mass of 10 mg will require total about $6\times10^4\,\mathrm{cm^2}$ surface area. The advantage of the modular design presented above is that the target mass can be scaled up relatively easily when needed. In the initial pilot phase, the experiment can begin with a smaller, e.g., $\sim \mathcal{O}(10\,\mu{\rm g})$ target mass. Afterward, it can be scaled up in a phased manner. In the last few years, graphene sheet manipulation and stacking have improved manyfold. It is now possible to transfer graphene from one substrate to another and even precisely control the stacking angle between the layers, if needed, at large scale\,\cite{Liu2022}. Moreover, the sensitivity of a BLG detector can be further enhanced by the use of photonic crystals and Fabry-P\'erot microcavities\,\cite{https://doi.org/10.1002/pssb.201800382,Efetov2018,doi:10.1021/nl302746n}. Especially, such devices can be tuned to enhance DM absorption rate in certain mass ranges. Lastly, we want to emphasize that graphene and other 2D materials form an active area of research with new applications being developed in photonics and twistronics. It is not too ambitious to imagine a scenario where some of these applications can be used to improve the detector design presented here. 


\subsection{Detector Noise \& Background}
The noise in such a device can be characterized by the noise equivalent power (NEP)\,\cite{10.1063/1.334129,5643118,https://doi.org/10.1002/adma.201704412}. The dark current from thermal excitation of the electrons from the valence to the conduction band may form a part of the noise. It can be estimated by calculating the overlap between the valence band electron distribution function at a temperature $T$ and the density of state in the conduction band. 
Notably, the BLG we propose offers the advantage of a tunable band gap, which can be adjusted, significantly reducing the dark current, as the dark current is proportional to $e^{-E_{\mathrm g}/k_B T}$. By operating the device at a low temperature, this noise can be mitigated.
Another component of the noise may come from the read-out circuit which will depend on the detailed design of the setup. As an example, we note that Ref.\,\cite{https://doi.org/10.1002/adma.201704412} reported NEP $\simeq 10^{-18}\,\mathrm{W\,Hz^{-1/2}}$ of a dual-gated BLG single photon detector operated at a much higher temperature of $100\K$. This value is on par with the existing superconductor-based photon detectors.

Apart from the internal noise, there can be additional environmental background sources of noise. Radioactive decays in the surroundings and the detector itself may be a source of background events similar to various other ongoing direct detection experiments. Such sources outside of the detector can be mitigated by proper shielding and creating radio-pure environment. A central volume inside the detector can also be fiducialized to further remove the background events from outside. Another important source of background events is the cosmic ray muons. The experiment can be placed in an underground laboratory to alleviate the cosmic ray-induced high energy background event rate. Optionally, an active veto can be implemented outside the detector (or integrated to the shielding box itself) to weed out the events originating from outside.


In this regard, we want to mention about an upcoming experiment called PTOLEMY that is planning to use tritium adsorbed onto a graphene substrate to detect low energy cosmic neutrinos\,\cite{Betti:2018bjv,Apponi:2021hdu,PTOLEMY:2019hkd,ptolemy_slides}. The technical design of a BLG-based DM detector will get help from PTOLEMY's prototype detector design and background study as their goal is also to achieve a large exposure ($\sim$ 100 g-yr for tritium, which implies even larger mass for graphene). Especially, the techniques that will be developed to navigate the challenges associated with scaling up graphene production, assembly, and subsequent processing will be directly helpful in developing a BLG-based detector.

\section{Conclusions \& Outlook}\label{sec:conclusion}
In this paper, we proposed bilayer graphene (BLG) as a target material for DM direct detection experiment and studied its sensitivity to probe DM-electron interaction. BLG is a 2D material consisting of two honeycomb lattices of carbon atoms stacked together. The electrons in a BLG exhibit a tunable band gap that can be adjusted with a gate voltage applied across the material. This band gap can be used as the energy threshold in the experiment giving us a convenient control on the thermal noise background. We computed the dielectric function of BLG using the random phase approximation and used it to compute the DM scattering and absorption rates. We then showed projections for limits on the DM-electron scattering cross section $\schie$ for a massless mediator, and on the kinetic mixing parameter $\kappa$ between dark photon DM and ordinary photon. Our results show that a 10 mg-yr experimental exposure with no background will allow us to probe $\schie\gtrsim10^{-38}\cmsq$ around $\mchi\simeq100\keV$, and reach the freeze-in benchmark models in the DM-electron scattering parameter space. Thanks to the small value of the band gap, BLG will be able to detect $\sim20\keV$ mass DM, and potentially even reach below $10\keV$ if lower energy threshold is achieved. Such small energy threshold may also allow us to probe the very low energy Earth-bound thermalized DM population predicted by some models. In the context of a dark photon model, DM absorption in BLG can probe $\kappa\gtrsim10^{-12}$ in the sub-eV mass range. We also computed the expected daily modulation in the DM scattering rate due to the rotation of the Earth about its axis. We found that the scattering rate can vary by more than the double of the average value for lighter DM. This modulation can help us distinguish the DM signal from other constant background sources.

The promising results from this study motivates us to ask next - how to build a DM detector using BLG? In this paper, we outlined a possible design of a detector inspired by some novel state-of-the-art techniques already achieved in graphene research. We found encouraging noise estimates for BLG that are competitive with conventional superconductor-based sensors. Going forward, we plan to pursue a dedicated experimental design study in a future work in collaboration with field experts. To this end, we also pointed out the PTOLEMY experiment which will offer technical help for scaling-up the detector.
In the projection calculation of this paper, we did not assume any background. However, in reality one should expect various sources of external backgrounds, such as cosmic rays, radioactive sources etc. The muons and gamma rays from the cosmic rays can be mitigated by doing the experiment in an underground laboratory. 
Appropriate shielding of the experimental setup and careful modeling of the radioactive sources will be needed. Additionally, there could be hardware-specific systematics that one might need to pay attention to\,\cite{Fuss:2022fxe}.

Both, direct detection of sub-MeV DM and graphene are relatively new areas of research with many novel ideas being developed today. Hence, there are plenty of scopes to improve the idea presented in this paper. For example, figure\,\ref{fig:gr_elf} clearly shows us that the sensitivity of bilayer graphene as DM target can be further improved by matching its band velocity $v$ with the DM velocity, or by extending the overlap between the detector response and DM phase space using flatter bands. Twisted bilayer\,\cite{doi:10.1073/pnas.1108174108, 2018Natur.556...43C, Cao2018Cor, Andrei2020} or multilayer ($>2$) graphene\,\cite{10.1143/PTPS.176.227, PhysRevB.78.045405, PhysRevB.77.155416} could be useful for this purpose. Recent research has seen significant progress in studying flat band materials, where the bands near the Fermi level exhibit flat characteristics, offering improvement in the sensitivity of DM detectors\,\cite{Regnault2022}. Various flat band materials, such as the kagome lattice, have been discovered\,\cite{Kang2020, Kang2020Dirac}. We want to pursue these possibilities in future.


\section*{Acknowledgment}
We thank Yoni Kahn, Noah Kurinsky, Anuvab Sarkar, and Rinchen Sherpa for helpful discussions during the work, and Ranjan Laha for suggestions to improve the manuscript. AD was supported by Grant Korea NRF-2019R1C1C1010050. 
J.J. and H.M. acknowledge support from the National Research Foundation of Korea (NRF) grants funded by the Korea government (MSIT) (Grants No. 2023R1A2C1005996) and the Creative-Pioneering Researchers Program through Seoul National University (SNU).

\bibliography{graphene}
\bibliographystyle{apsrev4-1}
\end{document}